\documentclass[journal]{IEEEtran}
\usepackage{marvosym}
\usepackage{textcomp}
\usepackage{xcolor}
\usepackage{graphicx}
\usepackage{amsmath}
\usepackage{amssymb}
\usepackage{amsfonts}
\usepackage{mathtools}
\usepackage{epsfig}
\usepackage{epstopdf}
\usepackage{hhline}
\usepackage{subfigure}
\usepackage{stfloats}
\usepackage{cite}
\usepackage{algorithm}
\usepackage{algorithmic}
\usepackage[CJKbookmarks=true,
bookmarksnumbered=true,
bookmarksopen=true]{hyperref}
\usepackage{multirow}
\usepackage{diagbox}

\begin{document}
\title{Efficient Deep Learning-based Cascaded Channel Feedback in RIS-Assisted Communications}
\author{Yiming~Cui, \emph{Graduate Student Member}, \emph{IEEE},~Jiajia~Guo, \emph{Member}, \emph{IEEE},

~Chao-Kai~Wen, \emph{Fellow}, \emph{IEEE}, and~Shi~Jin, \emph{Fellow}, \emph{IEEE}

\thanks{ 	
Y. Cui, J. Guo, and S. Jin are with the National Mobile Communications Research Laboratory, Southeast University, Nanjing 210096, China (e-mail: cuiyiming@seu.edu.cn; jiajiaguo@seu.edu.cn; jinshi@seu.edu.cn).
		
C.-K. Wen is with the Institute of Communications Engineering, National Sun Yat-sen University, Kaohsiung 80424, Taiwan (e-mail: chaokai.wen@mail.nsysu.edu.tw).

© 2024 IEEE.  Personal use of this material is permitted.  Permission from IEEE must be obtained for all other uses, in any current or future media, including reprinting/republishing this material for advertising or promotional purposes, creating new collective works, for resale or redistribution to servers or lists, or reuse of any copyrighted component of this work in other works.
}
}		

\maketitle

\begin{abstract}
In the realm of reconfigurable intelligent surface (RIS)-assisted communication systems, the connection between a base station (BS) and user equipment (UE) is formed by a cascaded channel, merging the BS-RIS and RIS-UE channels. Due to the fixed positioning of the BS and RIS and the mobility of UE, these two channels generally exhibit different time-varying characteristics, which are challenging to identify and exploit for feedback overhead reduction, given the separate channel estimation difficulty. To address this challenge, this letter introduces an innovative deep learning-based framework tailored for cascaded channel feedback, ingeniously capturing the intrinsic time variation in the cascaded channel.
When an entire cascaded channel has been sent to the BS, this framework advocates the feedback of an efficient representation of this variation within a subsequent period through an extraction-compression scheme. This scheme involves RIS unit-grained channel variation extraction, followed by autoencoder-based deep compression to enhance compactness. Numerical simulations confirm that this feedback framework significantly reduces both the feedback and computational burdens.
\end{abstract}


\section{Introduction}
Reconfigurable intelligent surfaces (RIS), consisting of a large number of tunable unit cells, are widely recognized as a promising technology in sixth-generation communications \cite{9530717}. By controlling the electromagnetic response of each unit, RIS can tailor the propagation environment to achieve better transmission \cite{8796365,9318531,ni2023analysis}. To realize high-quality RIS control and base station (BS) precoding/beamforming, the BS needs to acquire accurate downlink channel state information (CSI) \cite{10025776}.\footnote{RIS reflection pattern codebook-based implicit CSI acquisition is also a promising method for RIS-assisted communications, which is beyond the discussion scope of this paper. The interested readers can refer to \cite{an2024codebook} for further details.}  Given the lack of channel reciprocity in frequency division duplex (FDD) systems, a typical channel acquisition routine for FDD involves the user equipment (UE) first estimating the downlink CSI and then feeding the estimated downlink CSI back to the BS. However, the large number of unit cells in RIS results in extremely large dimensions of the BS-RIS-UE cascaded channel, leading to prohibitive feedback overhead. Developing an efficient feedback framework is essential to fully reap the benefits of RIS.

Numerous research works have been conducted to reduce channel feedback overhead in RIS-assisted communications. These works extend CSI feedback methods from massive multiple-input multiple-output (MIMO) systems and further incorporate the characteristics of RIS. A novel codebook-based feedback method, introduced in \cite{shen2021dimension}, utilizes channel sparsity in the angular domain to reduce feedback overhead. This method is further extended in \cite{shi2022triple} by exploiting the common information shared by the CSI of different UEs. In the past five years, deep learning (DL) has been introduced to CSI feedback, emerging as a popular topic in communications \cite{wen2018deep,guo2022overview}. DL-based CSI feedback, employing an autoencoder neural network, extracts environment knowledge from numerous CSI samples, achieving remarkable feedback performance with low computational complexity. DL-based CSI feedback was first introduced to RIS-assisted communications in \cite{xie2022quan}, adopting a novel Transformer-based neural network architecture to compress and reconstruct the BS-RIS-UE cascaded channel. Channel attention mechanism is also incorporated in neural architecture design to further improve the performance \cite{peng2024deep}.

Considering that the BS-RIS channel is relatively stable over a long period, a two-timescale DL-based CSI feedback method is proposed in \cite{guo2023deep,tang2023rismcnet} for RIS-assisted communications and achieves significant overhead reduction. In these works, the BS-RIS channel and the RIS-UE channel are separately compressed with two encoder networks, and fed back to the BS on each large and small timescale, respectively. However, this feedback framework is based on the separate channel estimation of the BS-RIS channel and the RIS-UE channel, which is difficult due to the passive characteristic of RIS. Although dual-link pilot transmission scheme \cite{hu2021two} can be used for the separate channel estimation, enabling novel channel acquisition design \cite{hu2021two,xu2022time}, this further requires the BS to operate in full-duplex mode. In most cases, the UE can only estimate the cascaded BS-RIS-UE channel in passive RIS-assisted communications. How to take advantage of the aforementioned characteristic in cascaded channel feedback still remains an unsolved challenge.

In this letter, an efficient DL-based cascaded channel feedback framework is proposed. We observe that though the stability of the BS-RIS channel is not explicit in the cascaded channel, it can be captured in each BS--RIS unit--UE component of the cascaded channel, manifested in its proportional variation. Therefore, we advocate the feedback of the cascaded channel on large timescales, and the feedback of the variation of the cascaded channel on a shorter timescale. An extraction-compression scheme can be employed to generate a compact representation of the variation for feedback, including RIS-unit-grained variation extraction and autoencoder-based deep compression. Simulation results show that the proposed method can effectively reduce feedback overhead and computational complexity compared with the DL-based feedback framework proposed in \cite{xie2022quan}.

\section{System Model}
\subsection{Signal Model}
\begin{figure}
    \centering
    \includegraphics[width=0.97\linewidth]{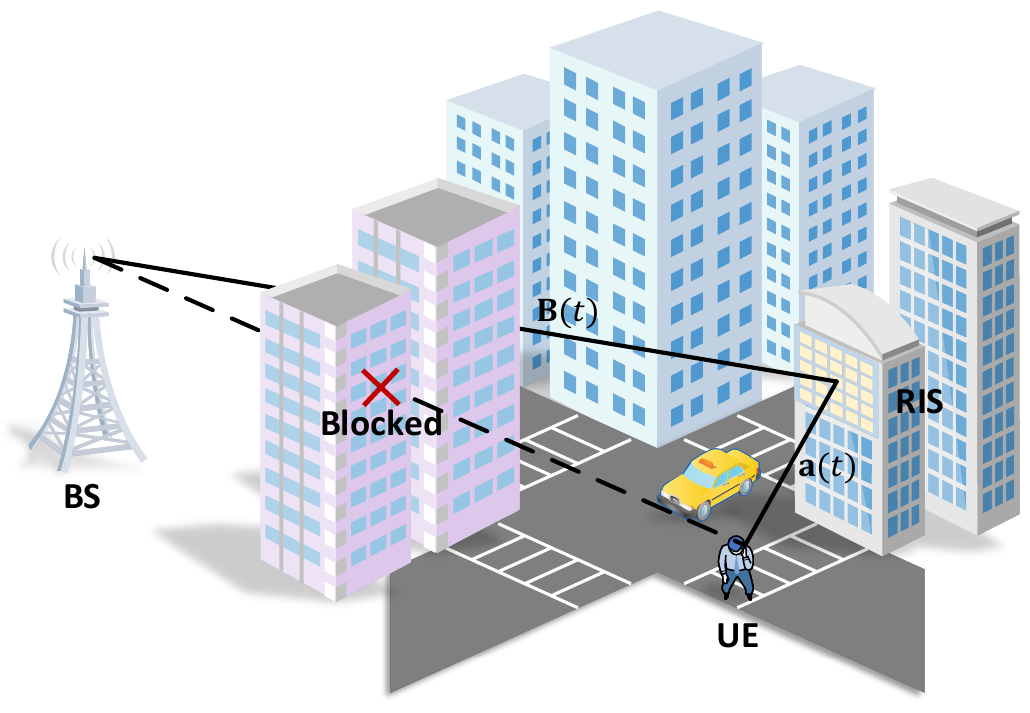}
    \caption{An example of RIS-assisted communication systems, where the direct link between the BS and the UE is blocked (unit: m)}
    \label{fig:ris}
\end{figure}

In this work, a RIS-assisted FDD system is considered, as illustrated in Fig. \ref{fig:ris}. The BS adopts a uniform linear array (ULA) with $N_{\rm t}$ antennas. The RIS uses a uniform planar array (UPA) with $N_{\rm RIS}=N_{\rm 1}\times N_{\rm 2}$ unit cells, where $N_{\rm 1}$ and $N_{\rm 2}$ are the row and column numbers of the UPA, respectively. The UE is equipped with a single antenna. The direct transmission link between the BS and UE is assumed to be completely blocked. The received signal at the $t$-th time interval can be formulated as follows:
\begin{equation}
    y(t)= \mathbf{a}^{\rm T}(t) \, {\sf diag}(\boldsymbol{\phi}(t)) \, \mathbf{B}(t) \mathbf{v}(t) x(t) + z(t),
\end{equation}
where ${\mathbf{a}(t)\in \mathbb{C}^{N_{\rm RIS}\times 1}}$, ${\boldsymbol{\phi}(t)\in \mathbb{C}^{N_{\rm RIS}\times 1}}$, ${\mathbf{B}(t)\in \mathbb{C}^{N_{\rm RIS}\times N_{\rm t}}}$, ${\mathbf{v}(t)\in \mathbb{C}^{N_{\rm t}\times 1}}$, ${x(t)\in \mathbb{C}}$, and ${z(t)\in \mathbb{C}}$ represent the RIS-UE channel, the reflection coefficients of RIS, the BS-RIS channel, the beamforming vector, the data-bearing symbol, and the complex additive Gaussian white noise, respectively. ${\sf diag}(\cdot)$ denotes the transformation from a vector to a diagonal matrix. By exploiting matrix properties, the received signal can be reformulated as:
\begin{equation}
    y(t)=\boldsymbol{\phi}^{\rm T}(t){\sf diag}(\mathbf{a}(t)) \mathbf{B}(t) \mathbf{v}(t) x(t) + z(t).
\label{equ:switch}
\end{equation}
We then define the effective cascaded BS-RIS-UE channel $\mathbf{H}(t)\in \mathbb{C}^{N_{\rm RIS}\times N_{\rm t}}$ as follows:
\begin{equation}
    \mathbf{H}(t)\triangleq{\sf diag}(\mathbf{a}(t)) \mathbf{B}(t).
    \label{equ:casc}
\end{equation}
From (\ref{equ:switch}), it is evident that acquiring the cascaded BS-RIS-UE channel $\mathbf{H}(t)$ is necessary to design the reflection coefficients $\boldsymbol{\phi}(t)$ and the beamforming vector $\mathbf{v}(t)$.

\subsection{DL-based CSI Feedback}

As a baseline, we briefly introduce the key framework of conventional DL-based CSI feedback, as outlined in \cite{xie2022quan}. The UE first compresses the cascaded channel $\mathbf{H}(t)$ into a low-dimensional codeword vector $\mathbf{s}(t)$ using an encoder neural network.\footnote{The cascaded channel is assumed to be perfect, i.e., channel estimation is not considered in this work.} This process can be formulated as:
\begin{equation}
    {\rm \mathbf{s}}(t)=\mathcal{Q}(f_\textrm{en}(\mathbf{H}(t);\Theta_\textrm{en})),
\label{equ:en}
\end{equation}
where $f_\textrm{en}(\cdot)$ denotes the encoder neural network (compression operation), $\Theta_\textrm{en}$ represents the neural network parameters of the encoder, and $\mathcal{Q}(\cdot)$ is the quantization function. $N_{\rm Q}$ bits are used to uniformly quantize each element of ${\rm \mathbf{s}}(t)$. Then, the codeword $\mathbf{s}(t)$ is fed back to the BS. Upon receiving the codeword, the BS uses a decoder neural network to reconstruct the cascaded channel from the codeword as:
\begin{equation}
    \mathbf{\widehat{\mathbf{H}}}(t)=f_\textrm{de}(\mathcal{D}({\rm \mathbf{s}}(t));\Theta_{\rm de}),
\label{equ:de}
\end{equation}
where $f_\textrm{de}(\cdot)$ represents the decoder neural network, $\Theta_\textrm{de}$ denotes the parameters of the decoder, and $\mathcal{D}(\cdot)$ is the dequantization function. To train this autoencoder network, the mean square error (MSE) is usually employed as a loss function, defined as:
\begin{equation}
    l_{\rm{MSE}}(\mathbf{H}(t),\widehat{\mathbf{H}}(t)) = \|\widehat{\mathbf{H}}(t)-\mathbf{H}(t)\|_{2}^{2},
\end{equation}
where $\| \cdot \|_2$ denotes the Euclidean norm.

\section{DL-based Two-Timescale Cascaded Channel Feedback}
In this section, key framework and neural network details of the proposed DL-based two-timescale cascaded channel feedback framework are introduced.

\subsection{Key Framework}
We propose a two-timescale feedback framework for the cascaded channel. For clarity, the cascaded channel
$\mathbf{H}(t)$ in (\ref{equ:casc}) can be rewritten as:
\begin{equation}
    \mathbf{H}(t) =
     \begin{bmatrix}
     a_1(t) \mathbf{b}_1(t)  \\
     \vdots \\
     a_{N_{\rm RIS}}(t) \mathbf{b}_{N_{\rm RIS}}(t)
    \end{bmatrix}  ,
\label{equ:ca}
\end{equation}
where $a_i(t)$ is the $i$-th element in the RIS-UE channel, $\mathbf{a}(t)$, and the vector, $\mathbf{b}_i(t)$, is the $i$-th row in the BS-RIS channel, $\mathbf{B}(t)$. The $i$-th row of the cascaded channel $\mathbf{H}(t)$ represents the BS--$i$-th RIS unit--UE cascaded channel, denoted as $\mathbf{h}_i(t)$.

\begin{figure*}[t]
	\centering
	\subfigure[\label{fig_first_case1} Two-timescale feedback protocol and illustration of the proposed two-timescale cascaded channel feedback framework]{\includegraphics[width=0.50\linewidth]{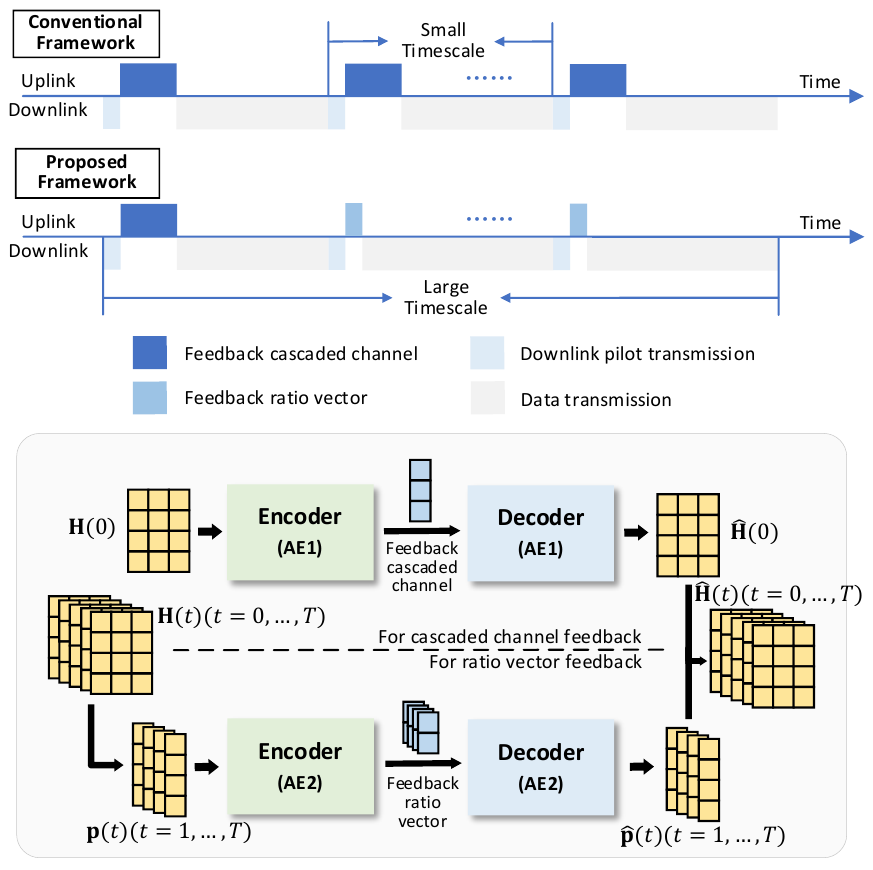}}
	\subfigure[\label{fig_second_case1} Architecture of AE1 and AE2]{\includegraphics[width=0.45\linewidth]{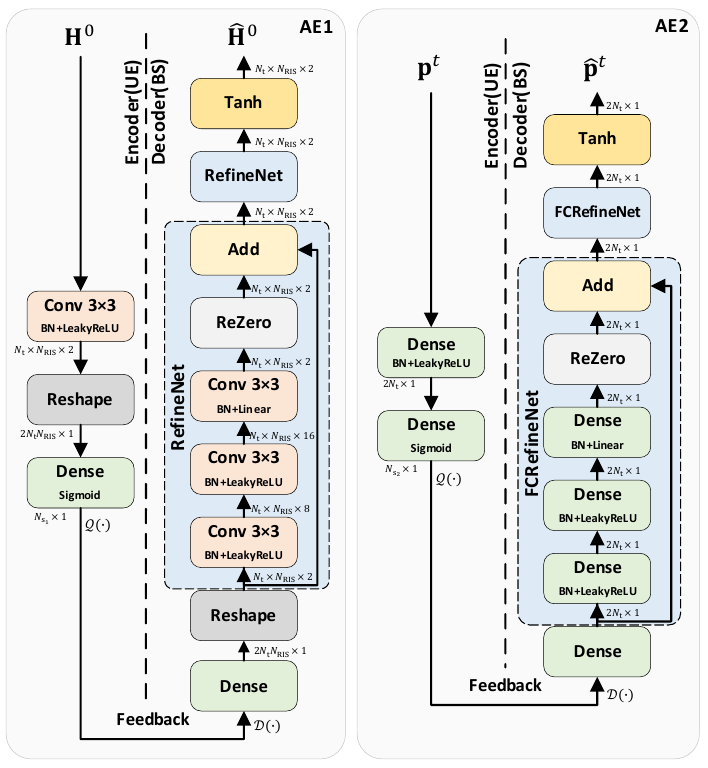}}
	\caption{Illustration of the proposed two-timescale cascaded channel feedback framework}
	\label{fig:nn}
	\vspace{-0.4cm}
\end{figure*}

\subsubsection{Variation Extraction and Compression}
Considering that the BS-RIS channel remains stable over a certain period \cite{guo2023deep}, that is, $\mathbf{b}_i(t) = \mathbf{b}_i(0)$, for $i=1, \ldots, N_{\rm RIS}$, the variation of the BS--$i$-th RIS unit--UE cascaded channel can be captured with a complex scalar. Hence, the variation of the whole cascaded channel can be represented by a ratio vector $\mathbf{p}(t)\in \mathbb{C}^{N_{\rm RIS}\times 1}$. The $i$th element of $\mathbf{p}(t)$ can be calculated as follows:
\begin{equation}
    p_i(t)=\frac{ \mathbf{h}_i^{\rm H}(0) \mathbf{h}_i(t)}{\| \mathbf{h}_i(0)\|_2^2 },
\label{equ:p}
\end{equation}
which in fact equals $a_i(t)/a_i(0)$.

After the extraction of the variation of the cascaded channel, we achieve a lower-dimension representation compared to the whole cascaded channel. However, the dimension of the ratio vector $\mathbf{p}(t)$ is still too large for feedback. Considering that the components of the ratio vector may also be correlated, we use an autoencoder network to further enhance the compactness of the ratio vector. This network is composed of an encoder network $f_\textrm{en,p}(\cdot)$ and a decoder network $f_\textrm{de,p}(\cdot)$, as follows:
\begin{subequations}
\begin{align}
    \mathbf{s}_\textrm{p}(t) &=\mathcal{Q}(f_\textrm{en,p}(\mathbf{p}(t);\Theta_\textrm{en,p})), \\
    \widehat{\mathbf{p}}(t) &=f_\textrm{de,p}(\mathcal{D}(\mathbf{s}_\textrm{p}(t));\Theta_\textrm{de,p}),
\end{align}
\end{subequations}
where $\mathbf{s}_\textrm{p}(t)$ is the codeword for the ratio vector feedback, $\widehat{\mathbf{p}}(t)$ is the reconstructed ratio vector at the BS, $\Theta_\textrm{en,p}$ and $\Theta_\textrm{de,p}$ are the neural network parameters of the encoder and the decoder, respectively. Consequently, suppose the cascaded channel $\mathbf{H}(0)$ has been fed back to the BS as $\widehat{\mathbf{H}}(0)$, each row of the cascaded channel $\widehat{\mathbf{H}}(t)$, for $i=1,\ldots,T$, can be reconstructed by
\begin{equation}
 \widehat{\mathbf{h}}_i(t) = \widehat{p}_i(t) \times \widehat{\mathbf{h}}_i(0).
\label{equ:p1}
\end{equation}

\subsubsection{Work Flow}

The two-timescale cascaded channel feedback method is then designed as follows. As shown in Fig. \ref{fig:nn}(a), two autoencoder networks, AE1 and AE2, are trained for cascaded channel feedback and ratio vector feedback, respectively. We assume that each large timescale and small timescale is composed of $T+1$ and $1$ time intervals, respectively, during which the BS-RIS and the RIS-UE channels remain stable. In the first time interval of each large timescale, the cascaded channel $\mathbf{H}(0)$ is fed back to the BS with AE1 and reconstructed as $\widehat{\mathbf{H}}(0)$. In the following $T$ time intervals, the UE calculates the ratio vector $\mathbf{p}(t)$ according to (\ref{equ:p}) and feeds back  $\mathbf{p}(t)$ to the BS with AE2. Once the BS reconstructs the ratio vector $\widehat{\mathbf{p}}(t)$, the cascaded channel $\widehat{\mathbf{H}}(t)$ can be acquired through (\ref{equ:p1}). The algorithmic description of the proposed method is shown in Algorithm \ref{alg:two}.

\begin{algorithm}[t]
	\caption{Two-Timescale Cascaded Channel Feedback}
	\label{alg:FEEL}
	\begin{algorithmic}[1]
		\REQUIRE Time intervals $T$, cascaded channel $\mathbf{H}(t)$ for $t=0,\ldots,T$, AE1, AE2.
		\STATE UE feeds back $\mathbf{H}(0)$ using AE1.
		\STATE BS reconstructs $\widehat{\mathbf{H}}(0)$.
            \FOR {$t=1,\ldots,T$}	
		    \STATE UE calculates $\mathbf{p}(t)$ per (\ref{equ:p}).
            \STATE UE feeds back $\mathbf{p}(t)$ using AE2.
            \STATE BS reconstructs $\widehat{\mathbf{p}}(t)$ and $\widehat{\mathbf{H}}(t)$ per \eqref{equ:p1}.
		\ENDFOR
	\end{algorithmic}
\label{alg:two}
\end{algorithm}

\subsection{Neural Network Details}

As illustrated in Fig. \ref{fig:nn}(b), two different autoencoder networks are used for cascaded channel and ratio vector feedback, respectively. The details are as follows:

For cascaded channel feedback, a convolutional autoencoder network, referred to as AE1, is employed to process the two-dimensional cascaded channel. The encoder consists of a convolutional layer and a dense layer to compress the cascaded channel into an $N_{\rm{s}_1}$-dimensional codeword. This codeword is then initially reconstructed with two RefineNet blocks, each primarily composed of three convolutional layers. The RefineNet block incorporates a residual structure \cite{he2016deep} and a ReZero module \cite{bachlechner2021rezero} to make the autoencoder networks easier to optimize. A Tanh function scales the output of the autoencoder network into the $(-1,1)$ range. Each convolutional layer is followed by a batch normalization layer, and LeakyReLU activation is employed for all convolutional layers except the last convolutional layer in each RefineNet.  For ratio vector feedback, a fully-connected autoencoder network, AE2, is used. AE2's architecture is similar to AE1's, including a two-layer structure to compress the ratio vector into an $N_{\rm{s}_2}$-dimensional codeword and a residual-structured decoder. The compression ratios of cascaded channel feedback and ratio vector feedback, denoted as $\gamma_1$ and $\gamma_2$, can be represented as follows:
\begin{equation}
    \gamma_1=\frac{N_{\rm{s}_1}}{2N_{\rm B}},~~\gamma_2=\frac{N_{\rm{s}_2}}{2N_{\rm RIS}}.
\end{equation}
Here, $N_{\rm B}=N_{\rm t}N_{\rm RIS}$ is the size of the cascaded channel.

In the proposed feedback framework, the cascaded channel is compressed and fed back to the BS with AE1 in each large timescale. This cascaded channel feedback is followed by $T$ times ratio vector feedback with AE2. The equivalent feedback overhead, $ N_{\rm equ}$, is formulated as:
\begin{equation}
 N_{\rm equ}=N_{\rm Q}\frac{  N_{\rm{s}_1}+  TN_{\rm{s}_2}  }{  T+1  }.
 \label{equ:overhead}
\end{equation}
Since the size of the cascaded channel $N_{\rm B}$ is typically significantly larger than $N_{\rm RIS}$,  especially in massive MIMO systems, this method can significantly reduce the feedback overhead compared to using AE1 to feed back the cascaded channel in each time interval.

\section{Simulation Results}
\subsection{Simulation Settings}

\subsubsection{Channel Generation Settings}
We use the Saleh-Valenzuela channel model to generate BS-RIS CSI samples \cite{dai2022distributed} and Wireless Insite to generate RIS-UE CSI samples \cite{wi}. The center frequency is set to 2.655 GHz, following the specifications of band ``n7'' in 3GPP TS 38.101-1 \cite{38101}. The BS is equipped with 32 antennas, while the RIS comprises $16 \times 16$ unit cells. The UE's position is randomly generated within a 30 m $\times$ 40 m rectangular area. The UE’s speed is set to 1 m/s, with a feedback interval of 5 ms. For the BS-RIS channel, we use the auto-regression model for time-varying channel generation employed in \cite{liu2022learning} with $\rho=0.9$. A total of 200,000 different cascaded channel samples are generated. The dataset is divided into training, validation, and test sets in an 8:1:1 ratio.

\subsubsection{Training Strategies and Neural Network Architectures}
All DL-based algorithms are implemented using TensorFlow 2.13.0 on a single NVIDIA Tesla V100 GPU. 4 bits are used to quantize each element of the codewords for both AE1 and AE2. An Adam optimizer is employed for neural network training. The learning rates for the two autoencoder networks are initially set to $10^{-3}$ and reduced to $10^{-4}$ when there is no further decrease in loss. The batch sizes are set to 64 for AE1 and 256 for AE2, respectively. MSE is adopted as the loss function. Normalized MSE (NMSE) is used to evaluate feedback performance, defined as \cite{guo2022overview}
\begin{equation}
	\rm NMSE= \rm E\left\{ {\|\mathbf{H}-\widehat{\mathbf{H}}\|_2^2}/{\|\mathbf{H}\|_2^2}\right\},
\end{equation}
where ${\rm E}\{\cdot\}$ represents expectation. The average NMSE over all $T+1$ intervals is adopted to evaluate the proposed method.

\subsubsection{Other Simulation Details}
For comparison purposes, we consider \cite{xie2022quan} as a baseline. Compared to the proposed method which flexibly selects the cascaded channel or the ratio vector for feedback in each time interval, this approach involves always feeding back the cascaded channel using autoencoder in each time interval. We simplify the Transformer-based autoencoder presented in \cite{xie2022quan} into a more lightweight autoencoder, which we name AE1. The equivalent feedback overhead of this baseline feedback method is denoted as $N_{\rm equ}=N_{\rm Q}N_{\rm{s}_1}$. \footnote{We also consider the typical RVQ codebook-based feedback method as a baseline with the same segmentation method as [7]. The NMSE performance is higher than 0 dB even when using 480 bits for feedback.}

\subsection{Performance Analysis}

\begin{figure}[t]
    \centering
    \includegraphics[width=0.45\textwidth]{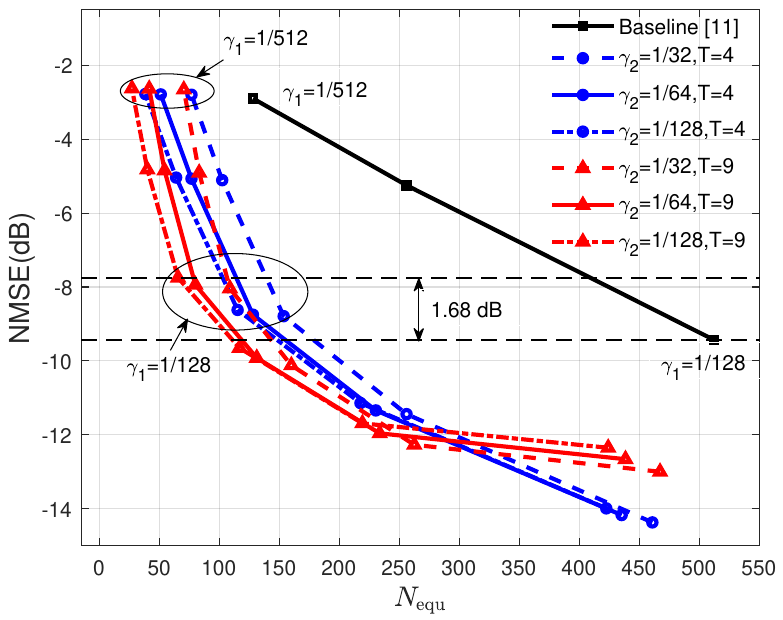}
    \caption{Performance of the DL-based two-timescale cascaded channel feedback method.}
    \label{fig:1}
\end{figure}

Fig. \ref{fig:1} summarizes the performance of the proposed two-timescale cascaded channel feedback method compared to the baseline method. The performance is evaluated with varying values of $T$, $\gamma_1$, and $\gamma_2$.
For each curve, $\gamma_2$ is fixed while the value of $\gamma_1$ is varied at intervals of $1/2$ as marked in Fig. \ref{fig:1}. When the same $\gamma_1$ is adopted, that is, the same AE1 is used for both the baseline and the proposed method, the proposed method exhibits a remarkable feedback overhead reduction compared to the baseline, with a performance drop less than 1.68 dB when feedback bits are not extremely large, i.e., less than 500. If the NMSE is required to be lower than $-9$ dB, the feedback overhead can be reduced by approximately 80\% when $T=9$. Additionally, when the codeword length is around 200, the proposed method achieves a performance gain of more than 7 dB compared to the baseline, demonstrating its effectiveness.

The influence of different hyperparameters is then investigated. Firstly, the feedback overhead can be further reduced as $T$ increases, with negligible loss in NMSE. The reduction in overhead is more evident when the feedback overhead is large, reaching close to 50\% when $T$ increases from 4 to 9. This indicates that a larger value of $T$ can be adopted within the channel coherence time to further reduce feedback overhead. Secondly, when $\gamma_1$ decreases, the NMSE significantly degrades for both the baseline and the proposed method, indicating that the feedback performance is strongly coupled with $\gamma_1$. This is because the performance in time interval $t$ is calculated based on the feedback performance of the cascaded channel in the initial time interval $(t=0)$, according to (\ref{equ:p1}). Finally, the performance is less sensitive to $\gamma_2$. Therefore, it can be concluded that the success of the proposed method is primarily attributed to the higher compressibility of the ratio vector $\mathbf{p}(t)$ compared to the cascaded channel. This is because the ratio vector $\mathbf{p}(t)$ represents the variation of the cascaded channel instead of the cascaded channel itself and further leverages the temporal correlation of the channel.

In addition, given the same feedback bit budget, maintaining an appropriate ratio between $\gamma_1$ and $\gamma_2$ can maximize the performance of the proposed method. This is because the overall feedback performance is determined by both the cascaded channel feedback at $T=0$ and the ratio vector feedback. Using either an excessively large $\gamma_1$ and a small $\gamma_2$, or an excessively large $\gamma_2$ and a small $\gamma_1$, will undermine the feedback performance with a fixed $N_{\rm equ}$.

\begin{figure}[t]
    \centering
  \includegraphics[width=0.45\textwidth]{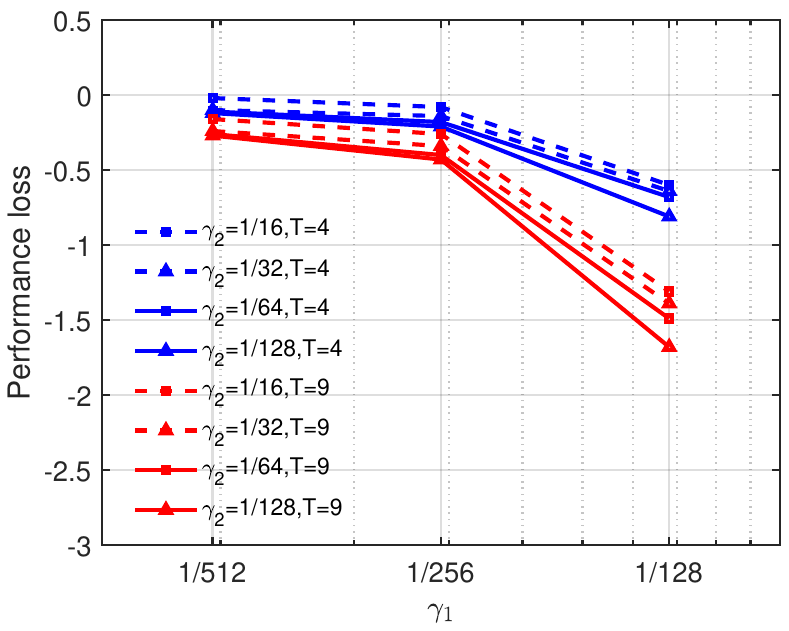}
    \caption{Performance loss caused by changing the last $T$ channel feedback from the whole cascaded channel feedback into the ratio vector feedback.}
    \label{fig:loss}
\end{figure}

Finally, we further investigate the performance loss caused by changing the last $T$ channel feedback from the whole cascaded channel feedback to the ratio vector feedback with different $\gamma_1$ and $\gamma_2$. The performance loss is shown in Fig. \ref{fig:loss}. We find that: 
\begin{enumerate}
\item The loss due to variation encoding increases with smaller $\gamma_2$.
\item The loss due to variation encoding decreases with smaller $\gamma_1$.
\end{enumerate}
Given that the error caused by ratio vector compression is fixed for a certain $\gamma_2$, the influence of the ratio vector compression error on overall feedback performance is more evident when the cascaded channel feedback at $t=0$ is accurate, i.e., when $\gamma_1$ is large. Conversely, when the cascaded channel feedback at $t=0$ is inaccurate, i.e., when $\gamma_1$ is small, the influence of the cascaded channel compression error becomes dominant, and the ratio vector compression error has little impact on overall feedback performance.

\subsection{Complexity Analysis}

\begin{figure}[t]
    \centering
    \includegraphics[width=0.45\textwidth]{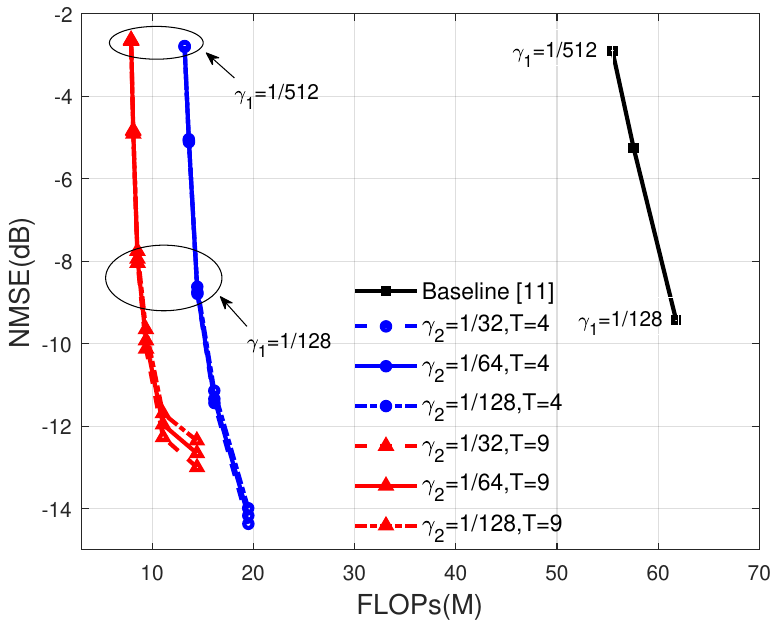}
    \caption{Computational complexity of the DL-based two-timescale cascaded channel feedback method.}
    \label{fig:3}
\end{figure}

The large dimension of the cascaded channel typically results in substantial computational complexity for DL-based CSI feedback. In the baseline method, frequent use of AE1 for cascaded channel feedback may lead to excessive computational overhead. Consequently, we analyze the computational complexity of the two autoencoder networks and compare the complexity between the proposed method and the baseline method. The Floating Point Operations (FLOPs) for the two different methods, across various compression ratios and values of $T$, are presented in Fig. \ref{fig:3}. Due to the high dimensionality of the cascaded channel, the FLOPs for AE1 are significantly higher than those for AE2. The computational complexity of the proposed method is approximately ${1}/{(T+1)}$ of the baseline method, yielding a significant reduction in computational overhead.

Furthermore, we compare the inference time of AE1 and AE2 using the computational resources mentioned in Section IV.A. In GPU settings, the inference times for AE1 and AE2 are 0.214 ms and 0.067 ms, respectively. In CPU settings, these times are 3.354 ms for AE1 and 0.102 ms for AE2. These results underscore that the proposed method is more computationally efficient than the baseline method.

\section{Conclusion}

In this paper, we propose a DL-based two-timescale cascaded channel feedback framework for RIS-assisted communications. This framework leverages the quasi-static characteristic of the BS-RIS channel to reduce feedback overhead. In each large timescale, the cascaded channel is fed back to the BS, while a ratio vector is transmitted on a shorter timescale. This vector comprises ratios between each row of the current cascaded channel and the cascaded channel most recently fed back, representing the variation in the cascaded channel. Two distinct autoencoder networks are employed for the feedback of the cascaded channel and the ratio vector, respectively. Simulation results demonstrate that our framework can significantly reduce both feedback overhead and computational complexity, especially when only the cascaded BS-RIS-UE channel is available.

\bibliographystyle{IEEEtran}
\bibliography{refer1}

\end{document}